\documentclass[draftcls,journal, onecolumn]{IEEEtran}
\usepackage{amssymb,amsmath,epsfig,graphicx,theorem, color}
\usepackage{amsfonts}
\usepackage{bm}
\newtheorem{Theo}{Theorem}
\newtheorem{Lem}{Lemma}

\usepackage{epsfig,rotating,setspace,latexsym,amsmath,epsf,amssymb,bm}
\usepackage{cite}

\IEEEoverridecommandlockouts

\IEEEoverridecommandlockouts

\title{Compressing Encrypted Data and Permutation Cipher}

\author{\IEEEauthorblockN{Wei Kang\\}
\IEEEauthorblockA{School of Information Science and Engineering\\
Southeast University\\ Nanjing, Jiangsu, P. R. China, 210096\\
\emph{wkang@seu.edu.cn \\}}\IEEEauthorblockN{Nan Liu\\}
\IEEEauthorblockA{National Mobile Communications Research Laboratory\\
Southeast University\\ Nanjing, Jiangsu, P. R. China, 210096\\
\emph{nanliu@seu.edu.cn }}
\thanks{This paper was presented in part at Annual Allerton Conference on Communications, Control and Computing, 2012.  This work is partially
supported by the National Basic Research Program of China
(973 Program 2012CB316004), the National Natural Science Foundation of China
under Grants $61271208$, $61201170$  and $61221002$,
the Research Fund of National Mobile Communications Research Laboratory,
Southeast University (No. 2014A02),
the Project-sponsored by SRF for ROCS, SEM and Qing Lan Project.}}
\begin{document}
%\date{}
\maketitle

\begin{abstract}
In a system that performs both encryption and lossy compression, the conventional way is to compress first and then encrypt the compressed data. This separation approach proves to be optimal. In certain applications where sensitive information should be protected as early as possible, it is preferable to perform encryption first and then compress the encrypted data, which leads to the concept of the reversed system. Johnson \emph{et al.} proposed an achievability scheme for the reversed system that has a modulo-sum encryption followed by a compression using Wyner-Ziv distributed source coding with side information. However, this reversed system performs worse than the conventional system in the sense that it requires more compression rate and secrecy key rate. In this paper, we propose a new achievability scheme for the reverse system where encryption is conducted by a permutation cipher and then the encrypted data is compressed using the optimal rate-distortion code. The proposed scheme can achieve the optimal compression rate and secret key rate, and therefore shows that reversing the order of encryption and compression does not necessarily compromise the performance of an encryption-compression system. The proposed system attains weak secrecy, and we show that the information leakage is mainly contributed by the type information of the sequence, which is not concealed by the permutation cipher. Given the type of the sequence, the rest of the information leakage vanishes exponentially.

%In this paper, we propose a reversed system of joint encryption and lossy compression by reversing the order of the encryption and compression blocks in the conventional system. We suggest to use permutation cipher to encrypt the i.i.d source  first and then compress the encrypted data using ordinary lossy compressor, which can achieve the optimal compression rate.  We then study the performance of the permutation cipher and prove that both the normal permutation cipher and the composite permutation cipher can achieve the optimal secret rate with  exponentially vanishing information leakage. In addition, we show that the design of the permutation cipher does not depend on the source distribution or the compression function (within a large set), which provides the separation of the encryption-decryption blocks and compression-reconstruction blocks. 
\end{abstract}

\newpage

\section{Introduction} \label{sec_intro}
\begin{figure*}%[h]
\centering
\includegraphics[width=6.5in]{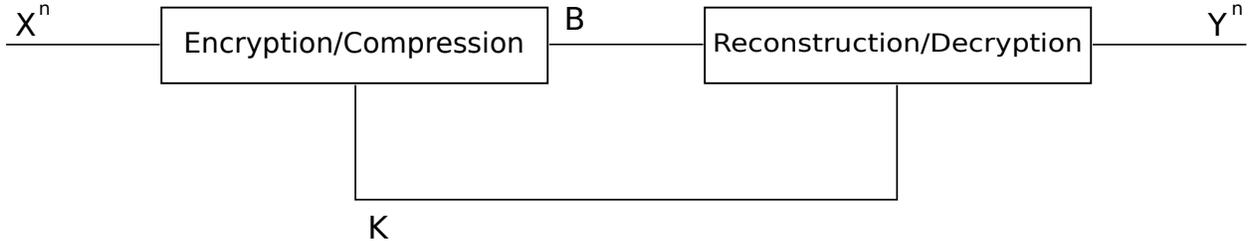}
\caption{The joint encryption and lossy compression system} \label{genesys}
\end{figure*}

%\subsection{Problem Formulation and Existing Result}
It is common for a communication system to incorporate both the encryption and lossy compression functions, see Fig. \ref{genesys}.  For example, distributing videos to authorized users via a public network requires an encryption on the videos to restrict the access from unauthorized users, and also compression of the videos to adapt to the traffic of the network.  

More formally, we consider an i.i.d. random sequence $X^n$ with distribution $P_X$ defined on a finite set $\mathcal{X}$. Assume a reconstruction set $\mathcal{Y}$ where the reconstruction sequence $Y^n$ is in $\mathcal{Y}^n$. Define a distortion function $d: \mathcal{X} \times \mathcal{Y} \mapsto \{0\} \bigcup \mathbb{R}^+$ and the distortion between $x^n\in\mathcal{X}^n$ and $y^n\in\mathcal{Y}^n$ is defined as 
\begin{align}
d(x^n,y^n)=\frac{1}{n} \sum_{i=1}^n d(x_i,y_i).
\label{nletter_dis}
\end{align}
Secret key is defined as a random variable $K$ uniformly distributed on $\{1,2,\dots, N\}$. 

A joint encryption-compression system is defined as follows.
\begin{align}
\bar{f}:&\mathcal{X}^n\times\{1,2,\dots,N\}\mapsto\{1,2,\dots,M\},\\
\bar{\phi}:&\{1,2,\dots,M\}\times\{1,2,\dots,N\}\mapsto\mathcal{Y}^n.
\end{align}

From the theory of rate distortion \cite{Csiszar:2011} and Shannon cipher \cite{Shannon:1949}, under the condition that the normalized information leakage $\frac{1}{n}I(X^n;\bar{f}(X^n,K))$ is sufficiently small, for a given distortion constraint $D$, an outer bound on the the compression and secret key rate pair $(R, R_s)\triangleq\left(\frac{1}{n}\log M, \frac{1}{n}\log N\right)$ in the joint encryption-compression system is
%\begin{figure*}[h]
%\centering
%\includegraphics[width=6.5in]{overall.eps}
%\caption{The joint encryption and lossy compression system} \label{genesys}
%\end{figure*}
%
\begin{align}
R&\ge \min_{P_{Y|X}:\textsf{E} d(X,Y)\le D} I(X;Y)+\epsilon,\label{lbound1}\\
R_s&\ge R+\epsilon,\label{lbound2}
\end{align}
for some arbitrarily small $\epsilon$.

\begin{figure*}
\centering
\includegraphics[width=6.5in]{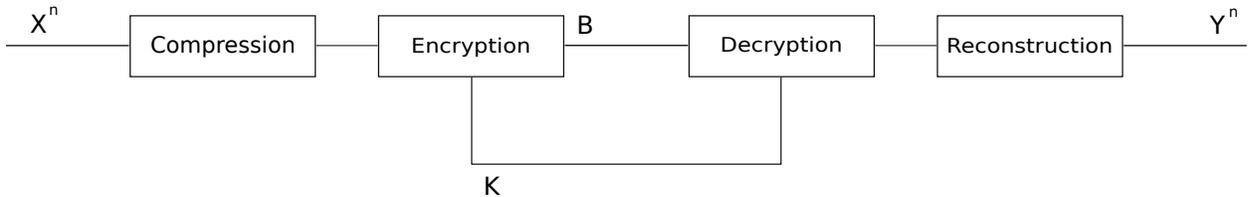}
\caption{The conventional system} \label{conventional}
\end{figure*}

The conventional system separates the functions of compression and encryption as illustrated in Fig. \ref{conventional}.  The system first compresses the source to the given rate, and then encrypts the compressed data with a secret key.  At the receiver side, it first decrypts the received data with the secret key and then performs a reconstruction of the source.
From classical results in information theory \cite{Csiszar:1981}, we know that the optimal rate distortion code achieves the bound in (\ref{lbound1}).
%\begin{align}
%R&=\min_{Ed(X,Y)\le D}I(X;Y)\label{rd}
%\end{align}
%and distortion between the source $X^n$ and the reconstruction $\varphi(f(X^n))$ satisfies
%\begin{align}
%E d(X^n,\varphi(f(X^n)))&\le D
%\end{align}
%where the distortion of a pair of sequences $(x^n,y^n)$ is defined as
%\begin{align}
%d(x^n,y^n)\triangleq \frac{1}{n}\sum_{i=1}^n d(x_i,y_i)
%\end{align}
Shannon's work \cite{Shannon:1949} shows that the cipher of modular addition achieves the bound in (\ref{lbound2}). Therefore, the conventional system is optimal.
%\begin{align}
%R_l\triangleq\frac{1}{n}I(B;g(B))=[R-R_{s}]^+
%\end{align}
%where $B=f(X^n)$ is the output from the compressor and the input of the cipher.
\begin{figure*}[t]
\centering
\includegraphics[width=6.5in]{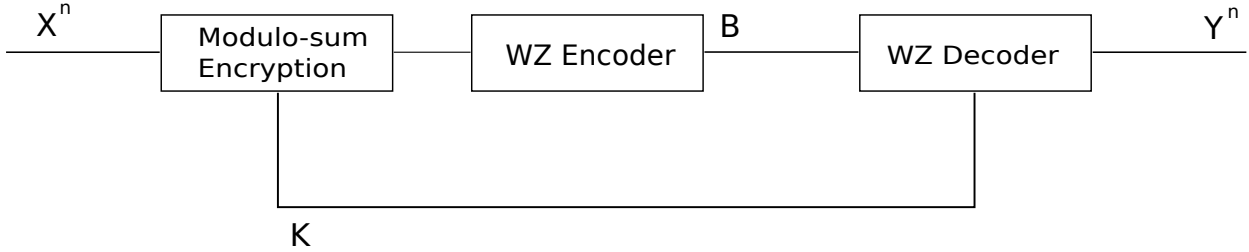}
\caption{The reversed system in \cite{Johnson:2004}} \label{half}
\end{figure*}
\begin{figure*}
\centering
\includegraphics[width=6.5in]{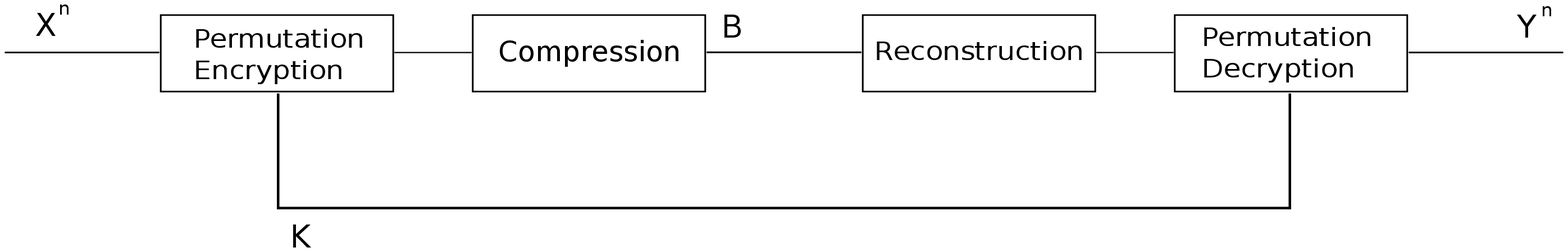}
\caption{The proposed reversed system with the permutation cipher} \label{reverse}
\end{figure*}

The properties of the conventional system are summarized as follows:
\begin{enumerate}
\item Optimality: it achieves the outer bounds in (\ref{lbound1}) and (\ref{lbound2}).
\item Perfect secrecy: the information leakage satisfies
\begin{align}
I(X^n;\bar{f}(X^n,K))=0.
\end{align}
\item Separation of blocks: the encryption and compression functions are separated at the encoder side, and the reconstruction and decryption functions are separated at the decoder side as well. 
\end{enumerate}

%\emph{comments: Shannon's work in [2] has any assumption on source i.i.d.? When we use Fig. 2, is the incoming data still i.i.d? Do we need to point this out? Also, the optimal rate-distortion code uses $(f, \varphi)$ but $f$ is used in your cipher as well. Need to fix this notation. Same with optimal cipher $(g,\phi)$}

In 2004, Johnson \emph{et al.}\ \cite{Johnson:2004} suggested that in certain applications, it may be preferable to perform compression \emph{after} encryption, i.e., reverse the order of the compression and encryption functions in Fig. \ref{conventional}. The main goal is to protect the sensitive source by encrypting as early as possible. For example, in the video distribution scenario \cite{Johnson:2004}, the video owner wants to perform the encryption to protect the video contents, but it may not have the incentive to conduct the compression.  The network operator, which is a different entity, desires to compress the data to adapt to the network traffic. The network operator may not be trusted by the contents distributor and therefore does not have the access to the secret key. In such a scenario, it is necessary to encrypt the source first and then compress the encrypted data. In \cite{Johnson:2004}, Johnson \emph {el al.}\ proposed a reversed system (see Fig. \ref{half}) where compression is performed after encryption.

In the achievability scheme for the reversed system proposed in \cite{Johnson:2004}, the source is first encrypted by a modulo-sum cipher with the help of a secret key, and then compressed by a Wyner-Ziv source encoder \cite{Wyner:1976}. At the receiver side, joint reconstruction-decryption is conducted by viewing the secret key as side information and performing the Wyner-Ziv decoding procedure. 

We note that the reversed system in \cite{Johnson:2004} has the following properties: 
\begin{enumerate}
\item Sub-optimality: it in general does not achieve the outer bounds in (\ref{lbound1}) and (\ref{lbound2}).
\item Weak secrecy: the information leakage in this reversed system is 
\begin{align}
\frac{1}{n}I(X^n;\bar{f}(X^n,K))\le \epsilon,\label{weaks}
\end{align}
for some arbitrarily small $\epsilon$, which is also the secrecy requirement in \cite{Wyner:1975}.
\item Semi-separation of the blocks: in this reversed system, the compression and the encryption functions are separated at the encoder side, but the reconstruction and decryption functions are performed jointly at the decoder side, because the secret key is viewed as the side information in the Wyner-Ziv decompression process.  
\end{enumerate}

In this paper, we propose a new achievability scheme for the reversed system. It consists of a permutation cipher followed by the optimal rate-distortion code. The proposed system achieves the same compression rate and secret key rate as the conventional system, i.e., (\ref{lbound1}) and (\ref{lbound2}), and therefore is optimal. As a result, we show that the performance of the joint compression-encryption system is not necessarily compromised when reversing the order of the blocks. 
We further study the performance of the permutation cipher. Unlike the modulo-sum cipher, the permutation cipher does not conceal the empirical distribution or ``type'' (see \cite{Csiszar:2011}) information of the source sequence $X^n$, which we denote as $P_{X^n}$. The type information is at the order of $\log n$ and the leakage of the type information does not violate the weak secrecy criterion as in (\ref{weaks}). We will show that  
%We show that, while the type information is not concealed by the permutation cipher, 
given the type of $X^n$, the information leakage of the proposed reversed system converges to zero exponentially.   

In summary,  the proposed reversed system based on permutation cipher has the following properties:
\begin{enumerate}
\item Optimality: it achieves the outer bounds in (\ref{lbound1}) and (\ref{lbound2}).
\item Weak secrecy: the type information is not concealed. But, given the type, the information leakage vanishes exponentially, i.e.,
\begin{align}
\frac{1}{n}I(X^n;\bar{f}(X^n,K)) \leq \frac{1}{n} H(P_{X^n})+\frac{1}{n}I(X^n;\bar{f}(X^n,K))|P_{X^n}), \label{leak01}
\end{align}
and
\begin{align}
H(P_{X^n}) & \leq  |\mathcal{X}|\log(n+1),\label{leak02}\\
I(X^n;\bar{f}(X^n,K))|P_{X^n}) &\le\exp(-n\mu). \label{leak03}
\end{align}
%and therefore the proposed system attains strong secrecy.
\item Separation of blocks: the encryption and compression functions are separated at the encoder side, and the reconstruction and decryption functions are separated at the decoder side as well. 
\end{enumerate}

The remainder of the paper is organized as follows. We propose the permutation cipher based reversed system in the next section. In section III, we will show that given the type,  the information leakage of the proposed system vanishes exponentially, which is followed by the conclusion.

\section{Reversed System Based on Permutation Cipher}
%\subsection{Reversed System Based on Permutation Cipher}
In this paper, we propose a reversed system consisting of a permutation cipher and a lossy compressor as in Fig. \ref{reverse}.  We define the encryption and decryption functions as
\begin{align}
f:\mathcal{X}^n\times\{1,2,\dots,N\}\mapsto\mathcal{X}^n,\\
\phi:\mathcal{Y}^n\times\{1,2,\dots,N\}\mapsto\mathcal{Y}^n,
\end{align}
and compression and reconstruction functions as
\begin{align}
g:\mathcal{X}^n\mapsto\{1,2,\dots,M\},\\
\varphi:\{1,2,\dots,M\}\mapsto \mathcal{Y}^n.
\end{align}
For encryption-decryption blocks, we have the following two kinds of permutation ciphers, which we refer to as type I and type II permutation cipher.
%{\color{red} Figure 4 should have permutation cipher in place of the encryption box and permutation decipher in place of the decryption box}
%\begin{figure*}
%\centering
%\includegraphics[width=6.5in]{reverse.eps}
%\caption{The proposed reversed system with the permutation cipher} \label{reverse}
%\end{figure*}

%{\color{red} I suggest you define the small blocks here, the $f$ and $g$}

\subsection{Type I Permutation Cipher}
The type I permutation cipher stores a group of permutations, and the value of the secret key is used to determine the specific permutation being used for the encryption. More specifically,  
assume $N$ elements from the symmetric group $\mathcal{S}_n$, namely $\left\{\pi_1,\pi_2,\dots,\pi_N\right\}\subset{\mathcal{S}_n}$, stored in the type I permutation cipher. Let $K$ denote the random key, which is uniformly distributed on $\{1,2,\dots,N\}$. The source $X^n$ is an i.i.d. sequence with length $n$ according to the distribution $P_X$. The encryption function of type I permutation cipher operates as
\begin{align}
f(X^n,K)=\pi_{K}(X^n), \label{encr}
\end{align} 
and the decryption function is 
\begin{align}
\phi(Y^n,K)=\pi_{K}^{-1}(Y^n). \label{decr}
\end{align}

\subsection{Type II Permutation Cipher}
One shortcoming of the type I permutation cipher we considered above is that exponentially many different permutations need to be stored in the both encryptor and decryptor, which is impractical. In 1982, Ahlswede and Dueck proposed in \cite{Ahlswede:1982} to construct a large group of permutations via the composition of a smaller number of permutations (also see \cite[Section 4.2]{Shulman:1995}). Inspired by \cite{Ahlswede:1982, Shulman:1995}, we consider the following permutation cipher, which we refer to as type II permutation cipher. We choose $L\triangleq\lceil\log N\rceil$ permutations from the symmetric group $\mathcal{S}_n$ and label them as $\sigma_1,\sigma_2,\dots, \sigma_L$. We then express the secret key in the form of a binary sequence as $K=(K_1,K_2,\dots,K_L)$, where $K_i\in\{0,1\}$ for $i=1,2,\dots,L$. We define the function 
\begin{align}
f_{K_i}&\triangleq\left\{\begin{array}{ll}
\sigma_i&\text{ if } K_i=1\\
\mathfrak{i}&\text{ if } K_i=0
\end{array}\right. \qquad i=1,2,\dots, L,
\end{align} 
where $\mathfrak{i}$ represents the identity mapping on $\{1,2,\dots,n\}$. The resulting permutation $\pi_K$ is defined as
\begin{align}
\pi_K\triangleq f_{K_L}\circ f_{K_{L-1}}\circ\cdots\circ f_{K_2}\circ f_{K_1},\label{compo}
\end{align}
where $\circ$ denotes the composition of the functions.
The encryption and decryption functions take on the same form as (\ref{encr}) and (\ref{decr}), with $\pi_K$ defined in (\ref{compo}).

Compared with the type I permutation cipher, which needs to store $N$ permutations, the type II permutation cipher only needs to store $L=\lceil\log N\rceil$ permutations. However, the type I permutation cipher only needs to perform one permutation operation at both the encryption and the decryption, while the type II permutation cipher may require up to $L$ permutation operations at both the encryption and the decryption. 

\subsection{Optimality of Compression Rate}
Whether a type I or type II permutation cipher is used, we always have that the output of the permutation cipher, i.e., $\pi_K(X^n)$, has the same distribution as the source $X^n$, since the source has an i.i.d. distribution. Thus, we can use a lossy compressor for the distribution $P_X$ as if the encryption/decryption pair does not exist. Rate-distortion theory guarantee that there exists a good lossy compressor achieving the outer bound in (\ref{lbound1}), which shows the optimality of the permutation cipher based reversed system regarding the rate of compression. We will show the optimality of the encryption, i.e., (\ref{lbound2}), in the next section.

\section{Information Leakage of the Permutation Cipher}
In this section, we prove that when the secret key rate satisfies (\ref{lbound2}), the information leakage of the permutation cipher based reverse system satisfies (\ref{leak01})-(\ref{leak03}), which shows that optimality of the encryption for the permutation cipher. 

%study the information leakage of the permutation cipher. More specifically, we consider feeding an i.i.d. source $X^n$ into a permutation cipher and then sending the output of the permutation cipher into a compression function $g$. We will present results regarding the  information leakage at the output of the function $g$ with respect to the source sequence $X^n$.

As we mentioned in the Section \ref{sec_intro}, the permutation cipher does not conceal the type information of the source sequence $X^n$. Actually, the type of $X^n$ should be provided to the compressor to achieve a satisfactory performance for the compression. 
We also note that the type information, i.e., $H(P_{X^n})$, is upper bounded by $|\mathcal{X}|\log(n+1)$ \cite{Csiszar:2011}, which is at the order of $\log n$ and the leakage of which does not violate the standard of weak secrecy. This proves (\ref{leak02}). Next, we will focus on the rest of the information leakage given the type information and prove (\ref{leak03}). We will follow a similar argument as in \cite[Chap. 17]{Csiszar:2011} and prove that given the type of the source sequence, the information leakage of the permutation cipher converges to zero exponentially.

We first consider a random permutation cipher, which represents the randomization among all the type I permutation ciphers, as follows. We assume $\Pi=\{\pi_1,\pi_2,\dots,\pi_N\}$ satisfies that $\pi_i$ is uniformly distributed in the symmetric group $\mathcal{S}_n$ for $i=1,2,\dots,N$ and $\pi_1,\pi_2,\dots,\pi_N$ are mutually independent. 

With respect to the randomization among all the type II permutation ciphers, we assume $L=\lceil\log N\rceil$ permutations $\sigma_1,\sigma_2,\dots,\sigma_L$ where $\sigma_i$ is uniformly distributed in the symmetric group $\mathcal{S}_n$ for $i=1,2,\dots,L$ and $\sigma_1,\sigma_2,\dots,\sigma_L$ are mutually independent. The resulting permutations $\Pi=\{\pi_1,\pi_2,\cdots,\pi_N\}$ as defined in (\ref{compo}) satisfy that $\pi_i$ is uniformly distributed in the symmetric group $\mathcal{S}_n$ for $i=1,2,\dots,N$ and $\pi_1,\pi_2,\dots,\pi_N$ are pairwise independent \cite[section 4.2]{Shulman:1995}. 

We note that the random type I and type  II permutation ciphers share the same marginal distribution. The difference is that the permutations in the random type I permutation cipher are mutually independent while the permutations in the  random type II permutation cipher are pairwise independent. %We will show that this difference will not cause huge performance gap with regard to the information leakage.

We consider a compression function $g:\mathcal{X}^n\mapsto\{1,2,\dots,M\}$. %We define set $g^{-1}(j)$ as
%\begin{align}
%g^{-1}(j)\triangleq \{x^n\in\mathcal{X}^n:g(x^n)=j\},\qquad j\in\{1,2,\dots,M\}
%\end{align}
Let us focus on a specific type $P$, which is not necessarily equal to $P_X$, i.e., the distribution of the i.i.d. source. 

\begin{Lem}\label{lbound}
The information leakage $I(X^n;g(\pi_K(X^n))|X^n\in\mathcal{T}_P^n,\Pi)$ over random permutation cipher $\Pi$ is upper bounded by
\begin{align}
I&(X^n;g(\pi_K(X^n))|X^n\in\mathcal{T}_P^n,\Pi)\le T_1+T_2+T_3,
\label{leakb}
\end{align}
where for some sufficiently large $\Delta>0$ and sufficiently small $\delta>0$, we have
\begin{align}
T_1&=\frac{M}{\Delta}\log\left|\mathcal{T}_P^n\right|,\\
T_2&=\left\{\begin{array}{ll}\log\left|\mathcal{T}_P^n\right|2\exp\left(-\frac{\delta^2}{2(2+\delta)}\frac{N}{\Delta}\right)&\text{for type I permutation cipher}\\
\log\left|\mathcal{T}_P^n\right|\left(\delta^2\frac{N}{\Delta}\right)^{-1}&\text{for type II permutation cipher}\end{array}
\right.,\\
T_3&=\delta.
\end{align}
%and $p$ will be specified in Lemma \ref{pbound}.
\end{Lem}
%\begin{proof}
The detailed proof of Lemma \ref{lbound} is given in Appendix \ref{lboundA}. Here we provide a brief overview of the proof.

We note that the information leakage with random permutation cipher and given type is as follows
\begin{align}
I&(X^n;g(\pi_{K}(X^n))|X^n\in\mathcal{T}_P^n,\Pi)\nonumber\\
&=H(X^n|X^n\in\mathcal{T}_P^n)-H(X^n|g(\pi_{K}(X^n)),X^n\in\mathcal{T}_P^n,\Pi).
\end{align} 
Given $X^n \in \mathcal{T}_P^n$, the source is uniformly distributed in the type $\mathcal{T}_P^n$, i.e.,
\begin{align}
\textsf{Pr}(X^n=x^n|X^n\in\mathcal{T}_P^n)=\frac{1}{\left|\mathcal{T}_P^n\right|},\qquad x^n\in\mathcal{T}_P^n.\label{xpro}
\end{align}
To make the information leakage small, essentially, we need to make the following conditional probability 
\begin{align}
\mathsf{Pr}(X^n=x^n|X^n\in\mathcal{T}_P^n,g(\pi_{K}(X^n))=j,\Pi=\{\pi_1,\dots,\pi_N\}),\label{cpr}
\end{align}
 close to uniform distribution $\frac{1}{\left|\mathcal{T}_P^n\right|}$. 
%To make the  conditional entropy term in (\ref{xpro1})  close to $\log\left|\mathcal{T}_P^n\right|$, it is equivalent to make the corresponding conditional probability close to uniform distribution. 

We consider the following set
\begin{align}
g^{-1}_P(j)\triangleq \{x^n\in\mathcal{T}_P^n:g(x^n)=j\},\qquad j\in\{1,2,\dots,M\}.
\end{align}
We pass the above set through the inverse of the permutations $\pi_1,\dots,\pi_N$ and obtain the following $N$ sets
\begin{align}
\pi^{-1}_1\left(g^{-1}_P(j)\right),\pi^{-1}_2\left(g^{-1}_P(j)\right)\dots,\pi^{-1}_N\left(g^{-1}_P(j)\right).\label{Nsets}
\end{align}
To make the conditional probability in (\ref{cpr}) close to the uniform distribution, we essentially need that the $N$ sets in (\ref{Nsets}) form a cover of the type $\mathcal{T}_P^n$ and every sequence in the type is covered by about the same number of sets in (\ref{Nsets}). In other word, the permutations $(\pi_1,\dots,\pi_N)$ are well spread.
However, the above goal will not be achieved if either of the following two events happens
\begin{enumerate}
\item The size of the set $g^{-1}(j)$ is too small, which will cause the union of the sets in (\ref{Nsets})  to be not large enough to cover the whole type. To identify these small sets, 
we use a threshold $\frac{1}{\Delta}$ in the sense that the size of a set $g^{-1}(j)$ with size less than $\frac{|\mathcal{T}_P^n|}{\Delta}$ is called a small set. This event contributes to the term $T_1$ in the upper bound of the information leakage in Lemma \ref{lbound}.
\item Permutations are not well spread such that the conditional distribution is not close to the uniform distribution, by which we mean 
\begin{align}
\left|\mathsf{Pr}\left(X^n=x^n\middle|X^n\in\mathcal{T}_P^n,\pi_{K}(X^n)\in g^{-1}_{P}(j),\Pi\right)-\frac{1}{\left|\mathcal{T}_P^n\right|}\right|>\frac{\delta}{\left|\mathcal{T}_P^n\right|}.
\end{align}
This event contributes to the term $T_2$ in the upper bound in Lemma \ref{lbound}. 
\end{enumerate}
When the above two events are not happening, the conditional distribution is close to the uniform distribution and the corresponding information leakage is represented by the term $T_3$ in Lemma \ref{lbound}.
%\end{proof}

Now we begin to evaluate the upper bound specified in Lemma \ref{lbound}. We assume $\epsilon=\frac{1}{n}\log\frac{N}{M}$, and let
\begin{align}
\Delta&=M\exp(\frac{1}{2}n\epsilon),\label{set1}\\
N&=\Delta\exp(\frac{1}{2}n\epsilon),\label{set2}\\
\delta&=\exp\left(-\frac{1}{6}n\epsilon\right).\label{set3}
\end{align}
We evaluate the information leakage given in Lemma \ref{lbound} for the type I permutation cipher as follows
\begin{align}
I&(X^n;g(\pi_K(X^n))|X^n\in\mathcal{T}_P^n,\Pi)\nonumber\\
&\le \left(\exp(-\frac{1}{2}n\epsilon)+2\exp\left(-\frac{1}{5}\exp\left(\frac{1}{6}n\epsilon\right)\right)\right)|\mathcal{X}|\log(n+1)+\exp\left(-\frac{1}{6}n\epsilon\right)\nonumber\\
&\le \exp\left(-\frac{1}{7}n\epsilon\right). \label{eqNan01}
\end{align}

Similarly, we can evaluation the information leakage for the type II permutation cipher as follows
\begin{align}
I&(X^n;g(\pi_K(X^n))|X^n\in\mathcal{T}_P^n,\Pi)\nonumber\\
&\le \left(\exp(-\frac{1}{2}n\epsilon)+\exp\left(-\frac{1}{6}n\epsilon\right)\right)|\mathcal{X}|\log(n+1)+\exp\left(-\frac{1}{6}n\epsilon\right)\nonumber\\
&\le \exp\left(-\frac{1}{7}n\epsilon\right). \label{eqNan02}
\end{align}

Next, we have 
\begin{align}
I(X^n;g(\pi_K(X^n))|P_{X^n})&=\sum_{P}\textsf{Pr}(P_{X^n}=P)I(X^n;g(\pi_K(X^n))|X^n\in\mathcal{T}_P^n)\nonumber\\
&\le\sum_{P}\textsf{Pr}(P_{X^n}=P)\exp\left(-\frac{1}{7}n\epsilon\right)\nonumber\\
&\le \exp\left(-\frac{1}{7}n\epsilon\right).
\end{align}
Therefore, we can conclude that there exists deterministic permutation ciphers, both type I and type II, with exponentially small information leakage. This result is summarized in the next theorem.

\begin{Theo} \label{theo1}
For any compression function $g$ with rate $R$, secrecy rate $R_s>R$,  and  sufficiently large $n$, there exist both type I and type II permutation ciphers with rate $R_s$ and $\mu>0$ such that the information leakage given type satisfies
\begin{align}
I&(X^n;g(\pi_K(X^n))|P_{X^n})\le\exp(-n\mu).
\end{align}
\end{Theo}
Theorem \ref{theo1} proves (\ref{leak03}) and concludes the proof of the optimality of encryption of the permutation cipher in our proposed reversed system.

\section{Conclusion}
In this paper, we proposed a reversed system of joint encryption and lossy compression by reversing the order of the encryption and compression blocks in the conventional system. We suggested to use the permutation cipher to encrypt the i.i.d source  first and then compress the encrypted data using an  ordinary lossy compressor. The proposed  reversed system based on permutation cipher can achieve the optimal compression rate and secret key rate, same as in the conventional compression-first-encryption-second system. It shows that reversing the order of encryption and compression does not necessarily lead to performance loss of an encryption-compression system.   We then studied the performance of the permutation cipher and proved that given the type, the information leakage of the permutation cipher vanishes  exponentially. 

\appendices

\section{Proof of Lemma \ref{lbound}}\label{lboundA}
For simplicity, we define short forms for the conditional distribution as follows
\begin{align}
P(x^n|\{\pi_1,\dots,\pi_N\},j)&\triangleq\mathsf{Pr}(X^n=x^n|X^n\in\mathcal{T}_P^n,\pi_{K}(X^n)\in g^{-1}_{P}(j),\Pi=\{\pi_1,\dots,\pi_N\}),\\
P(j)&\triangleq\mathsf{Pr}(\pi_{K}(X^n)\in g^{-1}_{P}(j)|X^n\in\mathcal{T}_P^n),\\
P(x^n|j)&\triangleq\mathsf{Pr}(X^n=x^n|X^n\in\mathcal{T}_P^n,\pi_{K}(X^n)\in g^{-1}_{P}(j)),\\
P(\{\pi_1,\dots,\pi_N\}|x^n,j)&\triangleq\mathsf{Pr}(\Pi=\{\pi_1,\dots,\pi_N\}|X^n\in\mathcal{T}_P^n,X^n=x^n,\pi_{K}(X^n)\in g^{-1}_{P}(j)).
\end{align}

We analyze the information leakage with a random permutation cipher and a given type as follows
\begin{align}
I&(X^n;g_P(\pi_{K}(X^n))|X^n\in\mathcal{T}_P^n,\Pi)\nonumber\\
&=H(X^n|X^n\in\mathcal{T}_P^n)-H(X^n|g_P(\pi_{K}(X^n)),X^n\in\mathcal{T}_P^n,\Pi)\nonumber\\
&=\log\left|\mathcal{T}_P^n\right|-H(X^n|g_P(\pi_{K}(X^n)),X^n\in\mathcal{T}_P^n,\Pi)\label{xpro2}\\
&=\sum_{j=1}^{M}P(j)\sum_{x^n\in\mathcal{T}_P^n}P(x^n|j)\sum_{\{\pi_1,\dots,\pi_N\}\subset \mathcal{S}^n}P(\{\pi_1,\dots,\pi_N\}|x^n,j)\left[\log\left|\mathcal{T}_P^n\right|-\log \frac{1}{P(x^n|\{\pi_1,\dots,\pi_N\},j)}\right],\label{xpro1}
%&=\log\left|\mathcal{T}_P^n\right|-\sum_{j=1}^M\sum_{\Pi=\{\pi_1,\}}\mathsf{Pr}(\pi_{K}(X^n)\in g^{-1}_P(j))H(X^n|g_P(\pi_{K}(X^n))=j,X^n\in\mathcal{T}_P^n,\Pi)
\end{align}
where (\ref{xpro2}) follows from (\ref{xpro}). 

We consider the first event, i.e.,~ the set $g^{-1}_{P}(j)$ is too small. We note that if the set $g^{-1}_{P}(j)$ is too small, the fact that the encrypted data $\pi_K(X^n)$ falls into the set $g^{-1}_{P}(j)$ will reveal quite some information about the source $X^n$.  
To identify these small sets, 
we use a threshold $\frac{1}{\Delta}$ in the sense that the size of a set $g^{-1}_{P}(j)$ with size less than $\frac{\mathcal{T}_P^n}{\Delta}$ is called a small set. More specifically,
for function $g$, type $P$, we define set $\mathcal{E}(g,P,\Delta)$ as
\begin{align}
\mathcal{E}(g,P,\Delta)\triangleq\left\{x^n\in g^{-1}_{P}(j):\frac{\left|g^{-1}_{P}(j)\right|}{\left|\mathcal{T}_P^n\right|}\ge \frac{1}{\Delta}, j\in\{1,2,\dots,M\}\right\},
\end{align}
and define the following quantity
\begin{align}
\eta(g,P,\Delta)=1-\frac{\left|\mathcal{E}(g,P,\Delta)\right|}{\left|\mathcal{T}_P^n\right|}.
\end{align}
Here, $\mathcal{E}(g,P,\Delta)$ represents the union of all the  ``normal'' sets in the type and $\eta(g,P,\Delta)$ represents the ratio of the small sets in the type to the whole type. We note that 
\begin{align}
\mathsf{Pr}(\pi_{K}(X^n)=x^n|X^n\in\mathcal{T}_P^n)=\frac{1}{\left|\mathcal{T}_P^n\right|},
\end{align}
therefore, we can interpret the quantity $\eta(g,P,\Delta)$ as
\begin{align}
\eta(g,P,\Delta)=\mathsf{Pr}(\pi_{K}(X^n)\notin \mathcal{E}(g,P,\Delta)|X^n\in\mathcal{T}_P^n).
\end{align}

We also note that 
\begin{align}
M=|g|\ge \frac{\eta(g,P,\Delta)\left|\mathcal{T}_P^n\right|}{\frac{\left|\mathcal{T}_P^n\right|}{\Delta}},
\end{align}
where the inequality follows from the fact that the number of all the sets $g^{-1}_{P}(j)$ is larger than the number of all the small sets $g^{-1}_{P}(j)$. The above inequality implies
\begin{align}
\mathsf{Pr}(\pi_{K}(X^n)\notin \mathcal{E}(g,P,\Delta)|X^n\in\mathcal{T}_P^n)=\eta(g,P,\Delta)\le\frac{M}{\Delta}.\label{MDpr}
\end{align}
Therefore, we upper bound the information leakage in (\ref{xpro1}) as follows
\begin{align}
I&(X^n;g_P(\pi_{K}(X^n))|X^n\in\mathcal{T}_P^n,\Pi)\nonumber\\
&= \sum_{j: g^{-1}_P(j)\nsubseteq\mathcal{E}(g,P,\Delta)}P(j)\sum_{x^n\in\mathcal{T}_P^n}P(x^n|j)\sum_{\{\pi_1,\dots,\pi_N\}\subset \mathcal{S}^n}P(\{\pi_1,\dots,\pi_N\}|x^n,j)\left[\log\left|\mathcal{T}_P^n\right|-\log \frac{1}{P(x^n|\{\pi_1,\dots,\pi_N\},j)}\right]+\nonumber\\
&+\sum_{j: g^{-1}_P(j)\subseteq\mathcal{E}(g,P,\Delta)}P(j)\sum_{x^n\in\mathcal{T}_P^n}P(x^n|j)\sum_{\{\pi_1,\dots,\pi_N\}\subset \mathcal{S}^n}P(\{\pi_1,\dots,\pi_N\}|x^n,j)\left[\log\left|\mathcal{T}_P^n\right|-\log \frac{1}{P(x^n|\{\pi_1,\dots,\pi_N\},j)}\right]\nonumber\\
& \le\frac{M}{\Delta}\log\left|\mathcal{T}_P^n\right|+%\max_{j: g^{-1}_P(j)\subseteq\mathcal{E}(g,P,\Delta)}
\sum_{x^n\in\mathcal{T}_P^n}P(x^n|j^\ast)\sum_{\{\pi_1,\dots,\pi_N\}\subset \mathcal{S}^n}P(\{\pi_1,\dots,\pi_N\}|x^n,j^\ast)\left[\log\left|\mathcal{T}_P^n\right|-\log \frac{1}{P(x^n|\{\pi_1,\dots,\pi_N\},j^\ast)}\right]\nonumber\\
& =T_1+%\max_{j: g^{-1}_P(j)\subseteq\mathcal{E}(g,P,\Delta)}
\sum_{x^n\in\mathcal{T}_P^n}P(x^n|j^\ast)\sum_{\{\pi_1,\dots,\pi_N\}\subset \mathcal{S}^n}P(\{\pi_1,\dots,\pi_N\}|x^n,j^\ast)\left[\log\left|\mathcal{T}_P^n\right|-\log \frac{1}{P(x^n|\{\pi_1,\dots,\pi_N\},j^\ast)}\right],
\end{align}
where 
\begin{align}
j^\ast\triangleq\arg\max_{j: g^{-1}_P(j)\subset\mathcal{E}(g,P,\Delta)}\sum_{x^n\in\mathcal{T}_P^n}P(x^n|j)\sum_{\{\pi_1,\dots,\pi_N\}\subset \mathcal{S}^n}P(\{\pi_1,\dots,\pi_N\}|x^n,j)\left[\log\left|\mathcal{T}_P^n\right|-\log \frac{1}{P(x^n|\{\pi_1,\dots,\pi_N\},j)}\right].
\end{align}
We consider the second event that permutations are not well spread such that the conditional distribution is not close to the uniform distribution. We have the following lemma
\begin{Lem}\label{pbound}
For function $g$, type $P$, and $\delta>0$, 
the random selected permutations $\Pi=\{\pi_1,\pi_2,\dots\pi_{N}\}$,  where $\pi_i$ is uniformly distributed over $\mathcal{S}_n$, satisfy
\begin{align}
\left|\mathsf{Pr}\left(X^n=x^n\middle|X^n\in\mathcal{T}_P^n,\pi_{K}(X^n)\in g^{-1}_{P}(j),\Pi\right)-\frac{1}{\left|\mathcal{T}_P^n\right|}\right|>\frac{\delta}{\left|\mathcal{T}_P^n\right|},\qquad \text{for } g^{-1}_{P}(j)\subseteq\mathcal{E}(g,P,\Delta),
\end{align}
\begin{enumerate}
\item 
with probability at most
\begin{align}
2\exp\left(-\frac{\delta^2}{2(2+\delta)}\frac{N}{\Delta}\right),\label{mupr}
\end{align}
if $\pi_1,\pi_2,\dots\pi_{N}$ are mutually independent.
\item 
with probability at most
\begin{align}
\left(\delta^2\frac{N}{\Delta}\right)^{-1},
\label{papr}
\end{align}
if $\pi_1,\pi_2,\dots\pi_{N}$ are pairwise independent.
\end{enumerate}

\end{Lem}
The proof of Lemma \ref{pbound} is in Appendix \ref{pboundA}.

\emph{Remark}: This lemma shows that for a random permutation cipher, the probability, that the conditional probability $\mathsf{Pr}\left(X^n=x^n\middle|X^n\in\mathcal{T}_P^n,\pi_{K}(X^n)\in g^{-1}_{P}(j),\Pi\right)$ is  not close to $\frac{1}{\left|\mathcal{T}_P^n\right|}$, converges to zero as long as the output of the permutation cipher falls into a ``normal'' set $g^{-1}_{P}(j)$. However, this probability in the mutually independent case decays exponentially while the probability in the pairwise independent case is polynomially small.
 
We then define following set for every $x^n\in\mathcal{T}_P^n$, $j\in \{1,2,\dots,M\}$ with $g^{-1}_{P}(j)\subseteq \mathcal{E}(g,P,\Delta)$, and $\delta>0$
\begin{align}
%\mathcal{S}(x^n,j,\delta)&\triangleq\left\{\{\pi_1,\dots,\pi_N\}\subset \mathcal{S}^n: \left|P(x^n|\{\pi_1,\dots,\pi_N\},j)-\frac{1}{\left|\mathcal{T}_P^n\right|}\right|\le\frac{\delta}{\left|\mathcal{T}_P^n\right|}\right\}\\
S(x^n,j,\delta)&\triangleq\left\{\{\pi_1,\dots,\pi_N\}\subset \mathcal{S}^n: \left|P(x^n|\{\pi_1,\dots,\pi_N\},j)-\frac{1}{\left|\mathcal{T}_P^n\right|}\right|>\frac{\delta}{\left|\mathcal{T}_P^n\right|}\right\},
\end{align}
where $\mathcal{S}(x^n,j,\delta)$ represents the set of permutations ciphers which has a conditional probability $P(x^n|\{\pi_1,\dots,\pi_N\},j)$ not close to $\frac{1}{\left|\mathcal{T}_P^n\right|}$. Then the above lemma implies 
\begin{align}
\mathsf{Pr}&(\Pi\in\mathcal{S}(x^n,j,\delta)|X^n=x^n,X^n\in\mathcal{T}_P^n,\pi_{K}(X^n)\in g^{-1}_{P}(j),g^{-1}_{P}(j)\subset\mathcal{E}(g,P,\Delta))\nonumber\\
&\le \left\{\begin{array}{ll}2\exp\left(-\frac{\delta^2}{2(2+\delta)}\frac{N}{\Delta}\right)&\text{for type I permutation cipher}\\
\left(\delta^2\frac{N}{\Delta}\right)^{-1}&\text{for type II permutation cipher}\end{array}\right..
\end{align} 
We then continue upper bounding the information leakage as follows
\begin{align}
I&(X^n;g_P(\pi_{K}(X^n))|X^n\in\mathcal{T}_P^n,\Pi)\nonumber\\
&\le T_1+\sum_{x^n\in\mathcal{T}_P^n}P(x^n|j^\ast)\sum_{\{\pi_1,\dots,\pi_N\}\in S(x^n,j^\ast,\delta)}P(\{\pi_1,\dots,\pi_N\}|x^n,j^\ast)\left[\log\left|\mathcal{T}_P^n\right|-\log \frac{1}{P(x^n|\{\pi_1,\dots,\pi_N\},j^\ast)}\right]\nonumber\\
&+\sum_{x^n\in\mathcal{T}_P^n}P(x^n|j^\ast)\sum_{\{\pi_1,\dots,\pi_N\}\notin S(x^n,j^\ast,\delta)}P(\{\pi_1,\dots,\pi_N\}|x^n,j^\ast)\left[\log\left|\mathcal{T}_P^n\right|-\log \frac{1}{P(x^n|\{\pi_1,\dots,\pi_N\},j^\ast)}\right]\nonumber\\
&\le T_1+T_2+\sum_{x^n\in\mathcal{T}_P^n}P(x^n|j^\ast)\sum_{\{\pi_1,\dots,\pi_N\}\notin S(x^n,j^\ast,\delta)}P(\{\pi_1,\dots,\pi_N\}|x^n,j^\ast)\left[\log\left|\mathcal{T}_P^n\right|-\log \frac{1}{P(x^n|\{\pi_1,\dots,\pi_N\},j^\ast)}\right].
\end{align}

We note that for $\{\pi_1,\dots,\pi_N\}\notin S(x^n,j,\delta)$, we have
\begin{align}
P(x^n|\{\pi_1,\dots,\pi_N\},j)\le \frac{1+\delta}{\left|\mathcal{T}_P^n\right|},
\end{align}
which leads to
\begin{align}
\log\left|\mathcal{T}_P^n\right|-\log \frac{1}{P(x^n|\{\pi_1,\dots,\pi_N\},j)}\le \log(1+\delta)\le \delta,
\end{align}
for a sufficiently small $\delta>0$. Then we have
\begin{align}
I&(X^n;g_P(\pi_{K}(X^n))|X^n\in\mathcal{T}_P^n,\Pi)\nonumber\\
&\le T_1+T_2+\sum_{x^n\in\mathcal{T}_P^n}P(x^n|j^\ast)\sum_{\{\pi_1,\dots,\pi_N\}\notin S(x^n,j^\ast,\delta)}P(\{\pi_1,\dots,\pi_N\}|x^n,j^\ast)\delta\nonumber\\
&\le T_1+T_2+\delta,
\end{align}
which concludes the proof.

\section{Proof of Lemma \ref{pbound}}\label{pboundA}
A permutation $\pi_k\in\mathcal{S}_n$  is a one-to-one mapping from $\mathcal{T}_P^n$ to $\mathcal{T}_P^n$. Thus we have 
\begin{align}
\textsf{Pr}(\pi_{k}(X^n)=x^n|X^n\in\mathcal{T}_P^n)=\frac{1}{\left|\mathcal{T}_P^n\right|},\qquad x^n\in\mathcal{T}_P^n,k\in\{1,2,\dots,N\},
\end{align}
and
\begin{align}
\textsf{Pr}(\pi_{k}(X^n)\in g^{-1}_{P}(j)|X^n\in\mathcal{T}_P^n)=\frac{\left|g^{-1}_{P}(j)\right|}{\left|\mathcal{T}_P^n\right|},\qquad j\in\{1,2,\dots,M\},k\in\{1,2,\dots,N\}.\label{gpro}
\end{align}

By Bayes' rule, we have
\begin{align}
\mathsf{Pr}&\left(X^n=x^n\middle|X^n\in\mathcal{T}_P^n,\pi_{K}(X^n)\in g^{-1}_{P}(j),\Pi\right)\nonumber\\
&=\mathsf{Pr}\left(\pi_{K}(X^n)\in g^{-1}_{P}(j)\middle|X^n=x^n,X^n\in\mathcal{T}_P^n,\Pi\right)\frac{\mathsf{Pr}\left(X^n=x^n\middle|X^n\in\mathcal{T}_P^n,\Pi\right)}{\mathsf{Pr}\left(\pi_{K}(X^n)\in g^{-1}_{P}(j)\middle|X^n\in\mathcal{T}_P^n,\Pi\right)}\nonumber\\
&=\mathsf{Pr}\left(\pi_{K}(X^n)\in g^{-1}_{P}(j)\middle|X^n=x^n,X^n\in\mathcal{T}_P^n,\Pi\right)\frac{1}{\left|g^{-1}_{P}(j)\right|},\label{bay}
\end{align}
where (\ref{bay}) is from (\ref{xpro}) and (\ref{gpro}).

For a given $i\in\{1,2,\dots,N\}, x^n\in\mathcal{T}_P^n$, we define the random variable $\chi(i)$ as
\begin{align}
\chi(i)\triangleq\left\{
\begin{array}{ll}
1& \pi_{i}(x^n)\in g^{-1}_{P}(j)\\
0&\pi_{i}(x^n)\notin g^{-1}_{P}(j)
\end{array}
\right.,
\end{align}
which is the function of the random permutation $\pi_{i}$ and 
\begin{align}
\mathsf{Pr}(\chi(i)=1)=\frac{\left|g^{-1}_{P}(j)\right|}{\left|\mathcal{T}_P^n\right|}.\label{bay1}
\end{align}
Then we have
\begin{align}
\mathsf{Pr}\left(\pi_{K}(X^n)\in g^{-1}_{P}(j)\middle|X^n=x^n,X^n\in\mathcal{T}_P^n,\Pi\right)=\frac{1}{N}\sum_{i=1}^N\chi(i), \label{bay2}
\end{align}
where the probability is with respect to the key random variable $K$, and it depends on the realization of the random cipher $\Pi$.
From (\ref{bay}) and (\ref{bay2}), we have
\begin{align}
\mathsf{Pr}\left(X^n=x^n\middle|X^n\in\mathcal{T}_P^n,\pi_{K}(X^n)\in g^{-1}_{P}(j),\Pi\right)=\sum_{i=1}^N\frac{\chi(i)}{N\left|g^{-1}_{P}(j)\right|}.
\end{align}
For mutually independent permutations, which translates to mutually independent $\chi(i)$ for $i\in\{1,2,\dots,N\}$, by applying Chernoff bound, we have for any $\beta>0$
\begin{align}
\mathsf{Pr}&\left(\mathsf{Pr}\left(X^n=x^n\middle|X^n\in\mathcal{T}_P^n,\pi_{K}(X^n)\in g^{-1}_{P}(j),\Pi\right)>\frac{1+\delta}{\left|\mathcal{T}_P^n\right|}\right)\nonumber\\
&=\mathsf{Pr}\left(\beta\frac{\left|\mathcal{T}_P^n\right|}{\left|g^{-1}_{P}(j)\right|\Delta}\sum_{i=1}^N\chi(i)>\beta(1+\delta)\frac{N}{\Delta}\right)\nonumber\\
&=\mathsf{Pr}\left(\exp\left(\beta\frac{\left|\mathcal{T}_P^n\right|}{\left|g^{-1}_{P}(j)\right|\Delta}\sum_{i=1}^N\chi(i)\right)>\exp\left(\beta(1+\delta)\frac{N}{\Delta}\right)\right)\nonumber\\
&\le\exp\left(-\beta(1+\delta)\frac{N}{\Delta}\right)\mathsf{E}\prod_{i=1}^N\exp\left(\beta\frac{\left|\mathcal{T}_P^n\right|}{\left|g^{-1}_{P}(j)\right|\Delta}\chi(i)\right)\label{chern1}\\
&=\exp\left(-\beta(1+\delta)\frac{N}{\Delta}\right)\prod_{i=1}^N\mathsf{E}\exp\left(\beta\frac{\left|\mathcal{T}_P^n\right|}{\left|g^{-1}_{P}(j)\right|\Delta}\chi(i)\right)\label{chern2}\\
&=\exp\left(-\beta(1+\delta)\frac{N}{\Delta}\right)\left(1+\frac{\left|g^{-1}_{P}(j)\right|}{\left|\mathcal{T}_P^n\right|}\left(\exp\left(\beta\frac{\left|\mathcal{T}_P^n\right|}{\left|g^{-1}_{P}(j)\right|\Delta}\right)-1\right)\right)^N,\label{chern3}
\end{align}
where (\ref{chern1}) is due to Markov's inequality, 
(\ref{chern2}) is due to the fact that $\chi(i)$ are mutually independent  for $i\in\{1,2,\dots,N\}$, and
(\ref{chern3}) is the calculation of the expectation with the probability from (\ref{bay1}).

We note 
\begin{align}
\exp\left(\beta\frac{\left|\mathcal{T}_P^n\right|}{\left|g^{-1}_{P}(j)\right|\Delta}\right)-1&=\sum_{k=1}^\infty\frac{\left(\beta\frac{\left|\mathcal{T}_P^n\right|}{\left|g^{-1}_{P}(j)\right|\Delta}\right)^k}{k!}\label{taylor1}\\
&\le \sum_{k=1}^\infty\left(\beta\frac{\left|\mathcal{T}_P^n\right|}{\left|g^{-1}_{P}(j)\right|\Delta}\right)^k\nonumber\\
&\le \left(\beta\frac{\left|\mathcal{T}_P^n\right|}{\left|g^{-1}_{P}(j)\right|\Delta}\right)\sum_{k=0}^\infty\left(\beta\frac{\left|\mathcal{T}_P^n\right|}{\left|g^{-1}_{P}(j)\right|\Delta}\right)^k\nonumber\\
&\le \frac{\beta\frac{\left|\mathcal{T}_P^n\right|}{\left|g^{-1}_{P}(j)\right|\Delta}}{1-\beta\frac{\left|\mathcal{T}_P^n\right|}{\left|g^{-1}_{P}(j)\right|\Delta}},\label{taylor2}
\end{align}
where (\ref{taylor1}) and (\ref{taylor2}) follows from Taylor expansion. 
Therefore, we have
\begin{align}
1+\frac{\left|g^{-1}_{P}(j)\right|}{\left|\mathcal{T}_P^n\right|}\left(\exp\left(\beta\frac{\left|\mathcal{T}_P^n\right|}{\left|g^{-1}_{P}(j)\right|\Delta}\right)-1\right)
&\le1+\frac{\frac{\beta}{\Delta}}{1-\beta\frac{\left|\mathcal{T}_P^n\right|}{\left|g^{-1}_{P}(j)\right|\Delta}}\nonumber\\
&\le 1+\frac{\frac{\beta}{\Delta}}{1-\beta}\label{gbig}\\
&\le \exp\left(\frac{\frac{\beta}{\Delta}}{1-\beta}\right)\nonumber\\
&=\exp\left(\frac{\beta}{\Delta}\left(1+\frac{\beta}{1-\beta}\right)\right),\label{chern4}
\end{align}
where (\ref{gbig}) is due to the condition that $g^{-1}_{P}(j)\subset \mathcal{E}(g,P,\Delta)$, which implies $\frac{\left|g^{-1}_{P}(j)\right|}{\left|\mathcal{T}_P^n\right|}\ge \frac{1}{\Delta}$.
By combining (\ref{chern3}) and (\ref{chern4}), and setting $\beta=\frac{\delta}{2+\delta}$, we have the following bound
\begin{align}
\mathsf{Pr}\left(\mathsf{Pr}\left(X^n=x^n\middle|X^n\in\mathcal{T}_P^n,\pi_{K}(X^n)\in g^{-1}_{P}(j),\Pi\right)>\frac{1+\delta}{\left|\mathcal{T}_P^n\right|}\right)&\le \exp\left(-\beta(\delta-\frac{\beta}{1-\beta})\frac{N}{\Delta}\right)\nonumber\\
&=\exp\left(-\frac{\delta^2}{2(2+\delta)}\frac{N}{\Delta}\right).
\end{align}
On the other hand, 
\begin{align}
\mathsf{Pr}&\left(\mathsf{Pr}\left(X^n=x^n\middle|X^n\in\mathcal{T}_P^n,\pi_{K}(X^n)\in g^{-1}_{P}(j),\Pi\right)<\frac{1-\delta}{\left|\mathcal{T}_P^n\right|}\right)\nonumber\\
&=\mathsf{Pr}\left(\exp\left(-\beta\frac{\left|\mathcal{T}_P^n\right|}{\left|g^{-1}_{P}(j)\right|\Delta}\sum_{i=1}^N\chi(i)\right)>\exp\left(-\beta(1-\delta)\frac{N}{\Delta}\right)\right)\nonumber\\
&\le \exp\left(\beta(1-\delta)\frac{N}{\Delta}\right)\prod_{i=1}^N\mathsf{E}\exp\left(-\beta\frac{\left|\mathcal{T}_P^n\right|}{\left|g^{-1}_{P}(j)\right|\Delta}\chi(i)\right)\nonumber\\
&= \exp\left(\beta(1-\delta)\frac{N}{\Delta}\right)\left(1+\frac{\left|g^{-1}_{P}(j)\right|}{\left|\mathcal{T}_P^n\right|}\left(\exp\left(-\beta\frac{\left|\mathcal{T}_P^n\right|}{\left|g^{-1}_{P}(j)\right|\Delta}\right)-1\right)\right)^N.\label{taylor51}
\end{align}
We note that
\begin{align}
\exp\left(-\beta\frac{\left|\mathcal{T}_P^n\right|}{\left|g^{-1}_{P}(j)\right|\Delta}\right)-1&=\sum_{k=1}^\infty\frac{\left(-\beta\frac{\left|\mathcal{T}_P^n\right|}{\left|g^{-1}_{P}(j)\right|\Delta}\right)^k}{k!}\nonumber\\
&\le-\beta\frac{\left|\mathcal{T}_P^n\right|}{\left|g^{-1}_{P}(j)\right|\Delta}\left(1-\frac{1}{2}\beta\frac{\left|\mathcal{T}_P^n\right|}{\left|g^{-1}_{P}(j)\right|\Delta}\right)\label{taylor3}\\
&\le-\beta\frac{\left|\mathcal{T}_P^n\right|}{\left|g^{-1}_{P}(j)\right|\Delta}\left(1-\frac{1}{2}\beta\right),\label{taylor31}
\end{align}
where the inequality in (\ref{taylor3}) is valid when $\beta\le 4$.
Therefore, we have
\begin{align}
1+\frac{\left|g^{-1}_{P}(j)\right|}{\left|\mathcal{T}_P^n\right|}\left(\exp\left(-\beta\frac{\left|\mathcal{T}_P^n\right|}{\left|g^{-1}_{P}(j)\right|\Delta}\right)-1\right)
&\le1-\frac{\beta}{\Delta}\left(1-\frac{1}{2}\beta\right)\label{taylor4}\\
&\le \exp\left(-\frac{\beta}{\Delta}\left(1-\frac{1}{2}\beta\right)\right),\label{taylor5}
\end{align}
where (\ref{taylor4}) is from (\ref{taylor31}), and the inequality in (\ref{taylor5}) is due to Taylor expansion and a sufficiently large $\Delta$.
By combining (\ref{taylor51}) and (\ref{taylor5}), and setting $\beta=\delta$, we obtain the following bound.
\begin{align}
\mathsf{Pr}\left(\mathsf{Pr}\left(X^n=x^n\middle|X^n\in\mathcal{T}_P^n,\pi_{K}(X^n)\in g^{-1}_{P}(j),\Pi\right)>\frac{1+\delta}{\left|\mathcal{T}_P^n\right|}\right)&\le \exp\left(-\beta(\delta-\frac{1}{2}\beta)\frac{N}{\Delta}\right)\nonumber\\
&=\exp\left(-\frac{\delta^2}{2}\frac{N}{\Delta}\right).
\end{align}
The above bounds together with the union bound complete the proof of the first assertion of the lemma.

For the pairwise independent permutations, which implies pairwise independent $\chi(i)$ for $i\in\{1,2,\dots,N\}$, we have
\begin{align}
\mathsf{Var}\left(\sum_{i=1}^N\chi(i)\right)=\sum_{i=1}^N\mathsf{Var}\left(\chi(i)\right)=N\frac{\left|g^{-1}_{P}(j)\right|}{\left|\mathcal{T}_P^n\right|}\left(1-\frac{\left|g^{-1}_{P}(j)\right|}{\left|\mathcal{T}_P^n\right|}\right).
\end{align}
We apply Chebyshev's inequality and obtain, 
\begin{align}
\mathsf{Pr}&\left(\left|\mathsf{Pr}\left(X^n=x^n\middle|X^n\in\mathcal{T}_P^n,\pi_{K}(X^n)\in g^{-1}_{P}(j),\Pi\right)-\frac{1}{\left|\mathcal{T}_P^n\right|}\right|>\frac{\delta}{\left|\mathcal{T}_P^n\right|}\right)\nonumber\\
&=\mathsf{Pr}\left(\left|\frac{\left|\mathcal{T}_P^n\right|}{\left|g^{-1}_{P}(j)\right|\Delta}\sum_{i=1}^N\chi(i)-\frac{N}{\Delta}\right|>\delta\frac{N}{\Delta}\right)\nonumber\\
&\le\frac{\mathsf{Var}\left(\frac{\left|\mathcal{T}_P^n\right|}{\left|g^{-1}_{P}(j)\right|\Delta}\sum_{i=1}^N\chi(i)\right)}{\left(\delta\frac{N}{\Delta}\right)^2}\nonumber\\
&=\frac{N\frac{\left|g^{-1}_{P}(j)\right|}{\left|\mathcal{T}_P^n\right|}\left(1-\frac{\left|g^{-1}_{P}(j)\right|}{\left|\mathcal{T}_P^n\right|}\right)\left(\frac{\left|\mathcal{T}_P^n\right|}{\left|g^{-1}_{P}(j)\right|\Delta}\right)^2}{\left(\delta\frac{N}{\Delta}\right)^2}\nonumber\\
&=\frac{\left(1-\frac{\left|g^{-1}_{P}(j)\right|}{\left|\mathcal{T}_P^n\right|}\right)\left(\frac{\left|\mathcal{T}_P^n\right|}{\left|g^{-1}_{P}(j)\right|\Delta}\right)}{\delta^2\frac{N}{\Delta}}\nonumber\\
%&\le \frac{\frac{\left|\mathcal{T}_P^n\right|}{\left|g^{-1}_{P}(j)\right|\Delta}}{\delta^2\frac{N}{\Delta}}\nonumber\\
&\le \left(\delta^2\frac{N}{\Delta}\right)^{-1},\label{taylor52}
\end{align}
where the inequality in (\ref{taylor52}) is due to the condition that $g^{-1}_{P}(j)\subset \mathcal{E}(g,P,\Delta)$, which implies $\frac{\left|g^{-1}_{P}(j)\right|}{\left|\mathcal{T}_P^n\right|}\ge \frac{1}{\Delta}$.

\bibliographystyle{unsrt}
\bibliography{/Users/wkang/Documents/Dropbox/writing/refphd}

\begin{thebibliography}{1}

\bibitem{Csiszar:2011}
I.~Csiszar and J.~Korner.
\newblock {\em Information Theory: Coding Theorems for Discrete Memoryless
  Systems}.
\newblock Cambridge University Press, 2nd edition, 2011.

\bibitem{Shannon:1949}
C.~E. Shannon.
\newblock Communication theory of secrecy systems.
\newblock {\em Bell Syst. Tech. J.}, 28:656--715, 1949.

\bibitem{Csiszar:1981}
I.~Csiszar and J.~Korner.
\newblock {\em Information Theory: Coding Theorems for Discrete Memoryless
  Systems}.
\newblock Academic Press, 1981.

\bibitem{Johnson:2004}
M.~Johnson, P.~Ishwar, P.~Prabhakaran, D.~Schonberg, and K.~Ramachandran.
\newblock On compressing encrypted data.
\newblock {\em IEEE Trans. on Signal Processing}, 52(10):992 -- 3006, Octorber
  2004.

\bibitem{Wyner:1976}
A.~D. Wyner and J.~Ziv.
\newblock The rate-distortion function for source coding with side information
  at the decoder.
\newblock {\em IEEE Trans. Inform. Theory}, 22(1):1--10, 1976.

\bibitem{Wyner:1975}
A.~D. Wyner.
\newblock The wire-tap channel.
\newblock {\em Bell Syst. Tech. J.}, 54:1355--1387, 1975.

\bibitem{Ahlswede:1982}
R.~Ahlswede and G.~Dueck.
\newblock Good codes can be produced by a few permutations.
\newblock {\em IEEE Trans. Inform. Theory}, 28(3)(3):430--443, 1982.

\bibitem{Shulman:1995}
N.~Shulman.
\newblock {\em Coding Theorems for Structured Code Families}.
\newblock Master thesis, Tel Aviv University, 1995.

\end{thebibliography}
\end{document}